\documentclass{article}
\usepackage[T1]{fontenc} 
\usepackage[utf8]{inputenc} 
\usepackage{ismir,amsmath,cite,url}
\usepackage{graphicx}
\usepackage{color}
\usepackage{algorithm}
\usepackage{algpseudocode}
\usepackage{lipsum}
\usepackage{tabularx}
\usepackage{lineno}
\usepackage{makecell}

\title{Text Conditioned Symbolic Drumbeat Generation\\ using Latent Diffusion Models}

\twoauthors
  {Pushkar Jajoria} {University of Galway \\ \tt 0009-0006-3789-5372}
  {James McDermott} {University of Galway \\ \tt 0000-0002-1402-6995}

\begin{document}

\maketitle
\begin{abstract}

This study introduces a text-conditioned approach to generating drumbeats with Latent Diffusion Models (LDMs). It uses informative conditioning text extracted from training data filenames. By pretraining a text and drumbeat encoder through contrastive learning within a multimodal network, aligned following CLIP, we align the modalities of text and music closely. Additionally, we examine an alternative text encoder based on multihot text encodings. Inspired by music’s multi-resolution nature, we propose a novel LSTM variant, MultiResolutionLSTM, designed to operate at various resolutions independently. In common with recent LDMs in the image space, it speeds up the generation process by running diffusion in a latent space provided by a pretrained unconditional autoencoder. 

We demonstrate the originality and variety of the generated drumbeats by measuring distance (both over binary pianorolls and in the latent space) versus the training dataset and among the generated drumbeats. We also assess the generated drumbeats through a listening test focused on questions of quality, aptness for the prompt text, and novelty. We show that the generated drumbeats are novel and apt to the prompt text, and comparable in quality to those created by human musicians.
\end{abstract}%

\section{Introduction}\label{sec:introduction}
\begin{figure*}
 \centerline{
 \includegraphics[width=0.7\linewidth]{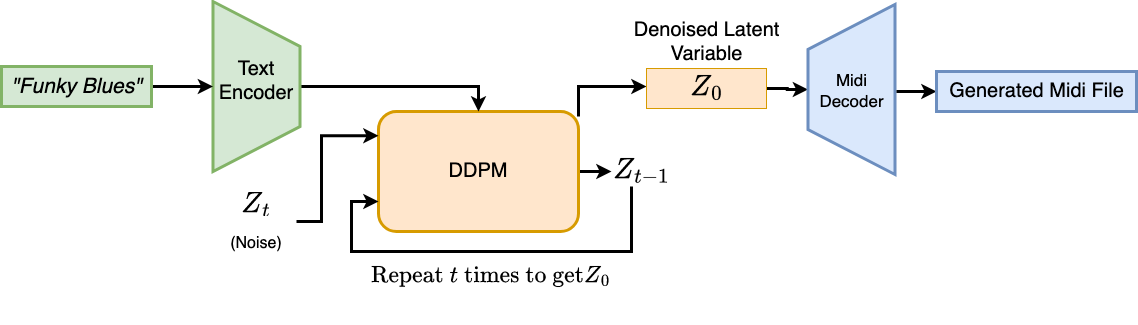}}
 \caption{Text conditioned MIDI file generation flow incorporating all elements of the model. The overall flow involves converting text prompts to text embeddings. These text embeddings along with noise ($Z_0$) are passed into a Latent Diffusion Model, and decoded to produce the final drumbeat. The color scheme used in this diagram -- Text Encoder in green and MIDI Decoder in blue -- is consistent throughout the paper.}
 \label{fig:generation_flow}
\end{figure*}

Research in AI-generated music has seen fast progress in recent years~\cite{biles2002genjam, eck2002first, roberts2018hierarchical, briot2019deep, Herremans_2017, Herremans_2019, musical_creativity_Hewitt, dahale2022generating}. Some recent deep learning models have successfully generated realistic-sounding music~\cite{huang2018music} by training on large datasets like \emph{The Lakh MIDI Dataset}. Although researchers have created models that can generate drum accompaniments (see Section~\ref{sec:related}), creating new drumbeats conditioned on text prompts remains to be solved. In this research we show that such a model can be trained by taking inspiration from state of the art text-conditioned {\em image} generation models~\cite{radford2021learning, ho2020denoising, nichol2021improved, rombach2022highresolution}.

Recent advances in Text-Conditioned Image Generation using diffusion models have laid the groundwork for generative systems, particularly when applied within latent spaces as opposed to the more natural pixel space~\cite{rombach2022highresolution} in the case of images. This shift to latent space has the benefit of dealing with a more compressed representation of data, which can lead to more efficient processing and the pretrained encoder-decoder can lead to potentially superior generation outcomes using the compressed representations. The adoption of diffusion techniques in latent space along with conditioning it on extraneous variables like text not only improves the model's stability during training but also enhances the quality and relevance. This paper builds upon these advantages, underscoring the potential latent space diffusion holds for generating musically coherent outputs that are responsive to textual prompts.

Conditioning the deep learning generative model on text in the case of music or drumbeat generation is not trivial. As opposed to the case of images, which have rich datasets linking text and images, music lacks such datasets. Commonly used MIDI datasets such as \emph{The Lakh Midi Dataset} and the {\em Magenta Groove MIDI Dataset} do not offer the rich text required to train such a model. To circumvent this issue, we work with the Groove Monkee dataset which provides descriptive filenames for its MIDI drumbeats (see Section~\ref{sec:dataset}). To deal with this text we investigate both a large language model (LLM) with CLIP-like pretraining, and a multihot keyword-based approach (see Section~\ref{sec:method}). Either approach gives a text encoding that can be used to guide drumbeat generation. Figure~\ref{fig:generation_flow} shows the overall architecture of our system. The provided text prompt is converted into text embeddings using the text encoder described in Section~\ref{subsec:text_encoder}. The random noise, denoted as $Z_{t_{\max}}$, along with these text encodings are autoregressively passed into the LDM that iteratively generates $Z_{t_{\max}} \dots Z_0$ is then passed into a pianoroll Decoder that generates the final pianoroll representing the drumbeat. This pianoroll is converted back into a MIDI file to get the final MIDI drumbeat.


In Section~\ref{sec:related}, we discuss some related research and how our research differs from it. We discuss the dataset in Section~\ref{sec:dataset}, including its text annotations and our preprocessing. We explain the three main parts of our model -- the text embedding layer, autoencoder, and diffusion -- in Section~\ref{sec:method}. Details of training are here also. This is followed by evaluation and results in Section~\ref{sec:eval_and_results}. Conclusions and future work are in Section~\ref{sec:conc}.

\section{Related Work}\label{sec:related}
We begin by highlighting the commonalities between drumbeat generation and image generation models using Latent Diffusion Models (LDM). LDMs have been used in the image domain with applications in the real world. Rombach et. al.\cite{rombach2022highresolution} have showcased generating high resolution images  using a similar architecture. The translation of these methods into music is not trivial. Firstly, music lacks such a rich dataset of text and symbolic music. Secondly, the data distribution of images and music is very different. Symbolic music in the form of a pianoroll is highly sparse. In addition to this, the relevant features in music are more temporal than local. The opposite is the case for images~\cite{MontesinosLópez2022}, motivating architectures like CNN. We solve both these issues by firstly, using a semi-labeled dataset (see Section~\ref{sec:dataset}). Secondly, we design a model based on RNNs as opposed to CNNs in order to better capture the temporal features in music. We take this a step further by creating a novel feature-extracting LSTM layer which works at multiple temporal resolutions. Thus we demonstrate that LDMs can work for a different data distribution other than images, i.e. drumbeats.

Research has shown that a good lower dimensional representation of drumbeats can be learned by a Variational Autoencoder (VAE), e.g.~a convolutional VAE~\cite{McDermott2021, Tikhonov}. We go beyond this research by introducing more recent deep learning architectures in the encoder and decoder and also conditioning the generation process on text. 

Researchers have also used methods other than AEs to generate novel drumbeats and drum accompaniments~\cite{dahale2022generating, makris2022conditional, wu2022jukedrummer, kaliakatsos2018generating, hoover2008scaffolding}. Both Kaliakatsos-Papakostas and Hoover and Stanley use evolutionary algorithms to generate drums: in one case to achieve conceptual blending, in the other for interactive control taking a ``scaffold'' from existing instrumental tracks. Dahale et al.~generate drumbeats as accompaniments by conditioning it on other instruments like \emph{string, bass, etc} using the Lakh Midi Dataset. Their model has no text conditioning element. Makris et al.~address the challenge of conditional drums generation by using a novel data encoding scheme inspired by Compound Word representation, focusing on a sequence-to-sequence architecture that utilizes a Bidirectional Long Short-Term Memory (BiLSTM) Encoder and a Transformer-based Decoder. There have been other experiments using MusicVAE~\cite{roberts2019hierarchical} to generate drumbeats by interpolating between known drumbeats. Beat Blender~\cite{Blankensmith2018} uses such a methodology to create an interactive website where users can do these interpolations on the fly. These methodology diverges from our research in the use of AEs and not conditioning the generation model on text or natural language. They also exclude note duration and velocity from the output, limiting expressivity. Other systems such as JukeDrummer\cite{wu2022jukedrummer} work with audio signals instead of symbolic music, often using VQ-VAE. We diverge from this research both from the perspective of output format and the underlying generation method. We also considered another annotated dataset of \emph{The Expanded Groove MIDI Dataset}~\cite{callender2020improving}. Although this dataset also offers annotations for each drumbeat in the form of genre tags and bpm, it does lack more information about the parts of the drumset that were used for creating that drumbeat. In addition to that we found the drumbeats in \emph{The Groove Monkee} dataset to be of high quality. Considering these two points, we decided on using \emph{The Groove Monkee} dataset for our research.

A recent publication by Wu et al.~have published a dataset of linked symbolic music and text along with using the linked dataset to train a contrastive learning model~\cite{wu2023clamp}. Since the dataset is not for drumbeats we opted to work with a different dataset instead. The authors have proved in this research that the text encoder learned by jointly training a multimodal music and text encoder is meaningful for downstream tasks and shows promising results. These results are consistent with image-text case as well~\cite{radford2021learning}. We take advantage of these learnings and extend the model architecture by using the generated text embeddings for generating text-conditioned drumbeats using LDMs.

\section{Dataset}\label{sec:dataset}

For the development and evaluation of our model, we use the Groove Monkee\footnote{See Ethics Statement.} dataset, a collection of MIDI drum loops. This dataset has a wide range of styles and genres like Rock, Blues, Latin, African, Electronic, etc., and song parts such as Verse, Chorus, and Fill. The dataset includes a total of 37,523 MIDI drum loops, including a variety of time signatures. The dataset is offered in a nested folder structure with each folder along with the MIDI file labeled accordingly. We use the 11,340 samples which are in simple time signatures. 


\subsection{MIDI Preprocessing}\label{subsec:midi_processing}

To use MIDI in a typical deep learning setting, we typically convert to a pianoroll format as follows. We follow a previously described lossy procedure~\cite{McDermott2021}. We extract metadata from the Groove Monkee MIDI files, such as the file's resolution, BPM, time signature, and track length in both beats and ticks. We assume that 128 time-slices are sufficient to represent 4 bars. Loops are tiled to give a standard length of 4 bars.  We assume 9 drum channels (kick, snare, closed hi-hat, open hi-hat, ride, crash, low-tom, mid-tom, high-tom) are sufficient to represent the large majority of tracks. All drum types (multiple kicks, snares, hi-hats, bongos, etc.) are mapped into one of the 9 channels. We create an array of $128 \times 9$ float values. A value of zero indicates no event, and a non-zero value indicates a note-on event. Note-off events are not represented but are rarely needed in drumbeats.

\subsection{Text Processing}

\begin{table}
\centering
\small
\begin{tabularx}{\columnwidth}{|X|}
\Xhline{2\arrayrulewidth}
\textbf{Sample Paths of MIDI files from the dataset} \\ 
\Xhline{2\arrayrulewidth}
Retro Funk GM/116 Say It/Fills/116 Say It Ride Fill 11.mid \\ \hline
RB GM/Straight Feel/100 Song 01 Straight/100 S1 Verse F6.mid \\ \hline
World Beats GM/Layered Beats/4-4 Layered Beats/145 Latin Rock 02.mid \\ \hline
Progressive GM/5-4 Grooves/180 5-4 02 F1.mid \\
\Xhline{2\arrayrulewidth}
\end{tabularx}
\caption{Example MIDI filepaths from the dataset.\label{tab:sample_paths}}
\end{table}

The textual metadata for our model was extracted from the hierarchical folder structure of the Groove Monkee dataset, which organizes MIDI files into folders and subfolders based on genre and other descriptive characteristics. Each MIDI file's filepath includes information indicative of its genre as well as specific attributes of the drumbeat it contains (see Table~\ref{tab:sample_paths} for examples). From this path, common identifiers such as \emph{``Groove Monkee"}, \emph{``GM"}, \emph{``Bonus"}, etc., were removed, resulting in a unique string for each MIDI file.
This approach allowed us to utilize the full path names as a proxy for the musical genre and characteristics of the drumbeats, under the assumption that the structured naming convention and folder organization provide a representative context for each MIDI file. 



\section{Method}\label{sec:method}
In this section, we discuss our methodology. The overall algorithm is as described in Algorithm~\ref{algo:training}.

We first train an AE model on drumbeat data, as described in Section~\ref{subsec:autoencoder}. The encoder maps from the data space to a latent space, and the decoder later maps back.

A large language model (LLM) text encoder which embeds the text information into text embeddings is described in Section~\ref{subsec:text_encoder}. We also hypothesized that due to the keyword-type text in the dataset, an alternative, training-free text encoder based on keyword multihot encoding could work as an alternative to the LLM, and this is described also. 

The DDPM model which runs the denoising process, guided by text, is described in Section~\ref{subsec:latent_diff}.

\begin{algorithm} 
\caption{Training Algorithm: Text-Conditioned MIDI Drumbeat Generation\label{algo:training}}
\begin{algorithmic}[1]
\State Create text encoder $T$ (either CLIP-like or multi-hot).
\State Train a pianoroll encoder and decoder $E$ and $D$ using reconstruction loss.
\For{each $(\text{pianoroll } m_i, \text{text } w_i)$ in Dataset}
    \State Get text embeddings $T(w_i)$.
    \State Get the latent vector $Z_0$ for $m_i$ using $E(m_i)$.
    \State Add noise to $Z_0$ to give $Z_1 \ldots Z_{t_{\max}}$.
    \State Train the DDPM model to predict $\epsilon_s := (Z_t - Z_{t-1})$, given $Z_t$, $T(w_i)$, and timestep $t$.
\EndFor
\end{algorithmic}
\end{algorithm}

\subsection{Text Encoding}\label{subsec:text_encoder}


The text information corresponding to each drumbeat is extracted by converting the path (including the filename) for each MIDI file into a string that represents that particular MIDI file. Given that the text is more oriented towards keywords rather than natural language, we chose to explore and compare two alternative methods for generating text embeddings. In Section~\ref{subsubsec:bert}, we discuss the text embeddings produced through contrastive Midi-Text pretraining, which involves adding a projection head over the pretrained BERT embeddings and then training the Midi and Text encoders using contrastive loss. The alternative method, focusing on keyword-based text embeddings, is examined in Section~\ref{subsubsec:mhv} where a multi-hot vector denoting the presence/absence of a curated list of musically relevant keywords that are present in the full dataset.

\subsubsection{Contrastive Language-MIDI Pretraining}\label{subsubsec:bert}
\begin{figure}
 \centerline{
 \includegraphics[width=1\columnwidth]{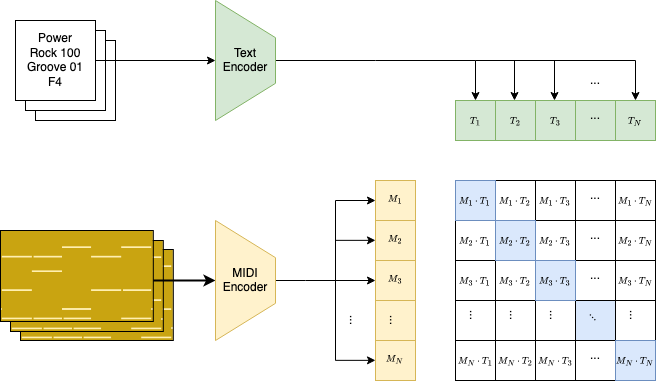}}
 \caption{Text supervised pretraining to train a text encoder in combination with a MIDI encoder to club both the text and drumbeat pianoroll together into a shared latent space, similar to CLIP~\cite{radford2021learning}. The MIDI encoder is discarded after training and only the text encoder is used. The text encoder consists of a projection head over \emph{BERT L-4 512} model which maps the 512 dimensional BERT embeddings into the final text embeddings.}
 \label{fig:clip}
\end{figure}

As is the case with CLIP\cite{radford2021learning}, we learn a multi-modal embedding space by jointly training a MIDI encoder and a text encoder. The model is trained to minimize the cosine similarity between $N^2 - N$ contrasting pairs $m_i$ and $t_j$ (\emph{i} $\ne$ \emph{j}), while maximizing the cosine similarity between $N$ same pairs $m_i$ and $t_i$. The loss function is a symmetric cross entropy loss over the similarity scores, i.e. both over $N$ MIDI embeddings given the text and $N$ text embeddings given the MIDI file. Figure~\ref{fig:clip} contains the overall architecture of this pretraining which is similar to the original paper except for a few key differences, i.e. we replace with image encoder with a MIDI encoder while making necessary changes in the encoder architecture to better capture the temporal dependencies of a drumbeat, which are absent in images. The text encoder contains a single projection head over the \emph{`bert\_uncased\_L-4\_H-512\_A-8'} embeddings. 

The trained text encoder is then used to generate the text embeddings that are provided as context to the denoising diffusion model discussed in Section~\ref{subsec:latent_diff}.

\subsubsection{Multihot text embedding}\label{subsubsec:mhv}
Since the nature of our dataset is more keyword-based and not natural language, we created a curated list of 57 keywords that were musically relevant in the top 95\% percentile of the keywords in the dataset. 
Similar to text encoder trained in the previous section, the denoising model ingests the textual information in the form of a Multihot Text Context Vector where a position in this vector is hot/active if that particular keyword is present in the text prompt. 
The limitation of this approach is that firstly, there is a limited number of tags that can be provided as text to the generation model; secondly, the model does not understand natural language and text prompts like “\emph{No Ride Cymbals}” would still yield in a drumbeat that has ride cymbals in them. We also append the bpm as an integer if present in the text prompt.


\subsection{Autoencoder}\label{subsec:autoencoder}

\begin{figure}
 \centerline{
 \includegraphics[width=1\columnwidth]{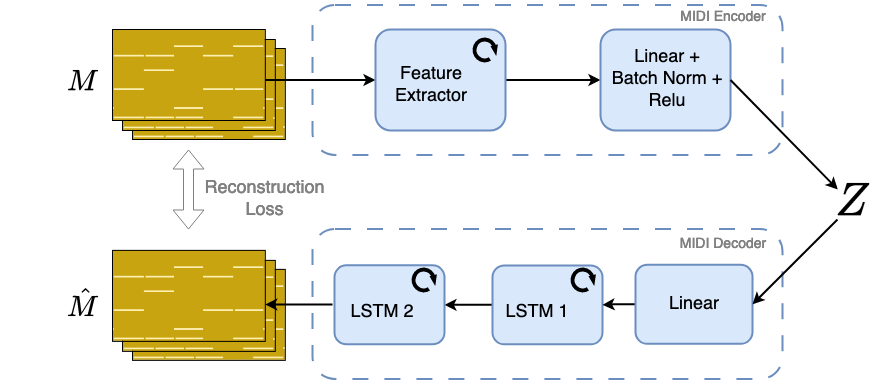}}
 \caption{We train an Autoencoder (AE) for pianoroll drumbeats using reconstruction loss. The trained MIDI encoder is used to generate the latent embeddings for a drumbeat pianoroll corresponding to a MIDI file that is needed to train a LDM. The MIDI decoder is used after the denoising process to generate a pianoroll drumbeat for a denoised $Z$. The MIDI encoder feature extractor consists of a 3-stacked LSTM\cite{6795963} which looks at the MIDI file at different resolutions. }
 \label{fig:autoencoder}
\end{figure}

For training an effective latent diffusion model, we need a MIDI encoder-decoder which can create a compact representation of a drumbeat. Just like a typical AE, the encoder transforms pianoroll drumbeats into a latent variable $Z$, which is then subsequently decoded back into the pianoroll space. The model is trained on reconstruction loss. The Figure~\ref{fig:autoencoder} describes the overall architecture of our AE. We also experimented with adding artificial noise to the encoded latent variable $Z$ in order to create a more robust latent space. The results of this are discussed in more detail in Section~\ref{sec:eval_and_results}. 

Motivated by the multi-resolution nature of music, we also introduce a novel Multi-Resolution LSTM (MRLSTM) component as part of the AE encoder. It is designed to analyze pianorolls at multiple temporal resolutions. In our implementation, the MRLSTM works at resolutions of 1:1, 1:2, and 1:4. The lowest resolution (1:4) focuses on every fourth time-slice, which in a typical simple-time rhythm will be of higher metric weight.


\subsection{Diffusion in Latent Space}\label{subsec:latent_diff}
To improve the stability and speed~\cite{rombach2022highresolution} of our diffusion process, we implement stable diffusion within a latent space. The  Latent Diffusion Model (LDM) is trained to learn the conditional probability distribution \begin{equation}
    p_{\theta}(\epsilon | Z_{t}, t, w)
\end{equation}
where $\epsilon$ is the noise present in the noised latent variable $Z_t$ after $t$ timestep of adding noise, with this process being conditioned upon textual information in the form of text embeddings $w$. This enables the iterative and autoregressive denoising of a randomly sampled $Z_t$, progressively estimating $Z_{t-1}$ through to $Z_0$, thus refining the generated output with each step. 

While at sampling time, $Z_t$ is sampled from a Normal Distribution, at training, we use the MIDI encoder discussed in Section~\ref{subsec:autoencoder} to generate $Z$. This latent embedding is then noised as per the noising schedule and by sampling a $t$ between 1-1000 to create $Z_t$. The loss for each step is computed using the mean squared error between $\epsilon$ and the predicted noise, $\hat{\epsilon}$. Algorithm 1 contains the training algorithm for reference.


\subsection{Model \& Training Details}\label{subsec:training}
The denoising model consists of 3 linear layers with batch normalization and ReLU activation. The input to the first linear layer is the concatenation of Z along with sinusoidal positional encoding of the timestep `t' and the text embeddings (either using the Multihot text embeddings or the text embeddings as per Section~\ref{subsubsec:bert}). The latent encoding is of 128 dimensions. We also pass an empty text embedding for 5\% of training steps to encourage the model to predict the noise without the text embeddings. This is especially useful when there are no active musical tags in the text prompt as per our curated list. 

\section{Experiments and Results}\label{sec:eval_and_results}


In this section, we describe how we tested our drumbeat generation deep learning model for creating MIDI drumbeats. 

Firstly, following~\cite{yang2020evaluation} we have compared distributions of inter-set and intra-set distances in different setups, visualised as probability density functions with kernel smoothing. We measure how dissimilar the generated drumbeats are, both from other generated drumbeats for the same text prompt (intra-set) and also from the most similar elements of the dataset (inter-set). 

Additionally, we experimented by adding random noise to the latent variable \(Z\) during training, to see how it affects these distances.

Lastly, we evaluated the quality of the generated music through a listening test. For this case, we avoid the problems associated with ``musical Turing tests''~\cite{yang2020evaluation} by focusing on questions of quality, suitability for the text, and novelty, rather than questions of artificiality.

\subsection{Empirical Experiments}\label{subsec:experiment}

We create 8 text prompts namely, \emph{`latin triplet', `4-4 electronic', `funky 16th', `rock fill 8th', `blues shuffle', `pop ride', `funky blues', `latin rock'} combining different genres and elements of a drumbeat. For each of these text prompts, 10 drumbeats are generated using our model. We compute the 45 pairwise distances using each of two metrics: Hamming Distance on the binarized pianorolls, and Euclidean Distance in the AE embedding space. In total we get 360 pairwise distance values for each metric. We also compute the distance of the generated drumbeats from the closest 1\% of the drumbeats from the dataset for each text-prompt variant. We first show these results in Table~\ref{tab:distance_metrics}. Comparing the third column with the first two, we see that the generated drumbeats are not mere repetitions of those in the dataset, i.e.~the model is generalising.

We also study the effects of text-conditioning by analyzing the Hamming distances between drumbeats generated from the same text prompts compared to those generated from different text prompts. Notably, the final column of Table~\ref{tab:distance_metrics} reveals a crucial insight: even when utilizing identical text prompts to generate multiple drumbeats, the resulting drumbeats exhibit differences. This variation highlights the variability inherent in the text-conditioned drumbeat generation model.

\begin{figure}
 \centerline{
 \includegraphics[width=1\columnwidth]{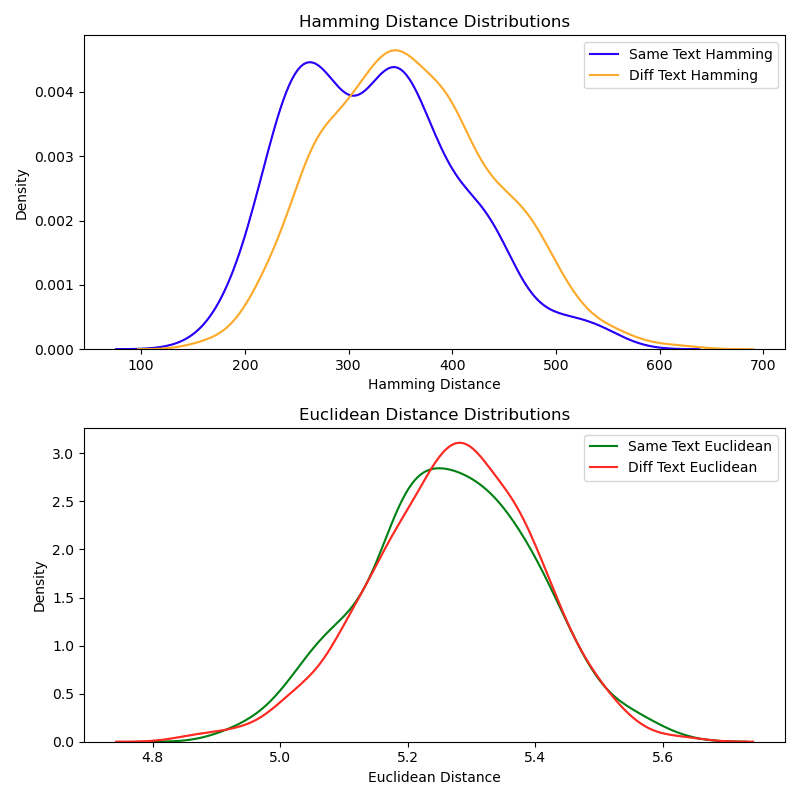}}
\caption{Density plots comparing Hamming (top) and Euclidean (bottom) distances of drumbeats generated from identical versus different text prompts.}
\label{fig:text_conditioning}
\end{figure}

Despite this variability, our further analysis, illustrated in Figure~\ref{fig:text_conditioning}, demonstrates a discernible impact of text-conditioning on drumbeat generation while using the Hamming distance. In the Euclidean space, while the same-text drumbeats do not cluster as closely as one might anticipate, given the autoencoder's design and training objective focused on just reconstruction loss, there remains a substantial difference between the distances of same-text and different-text drumbeats. This outcome suggests that, although the autoencoder does not explicitly encode genre-specific characteristics or ensure close clustering of similar genres in the latent space, text-conditioning nonetheless exerts a subtle but significant influence on the generated drumbeats.

\begin{table*}
    \centering
    \footnotesize
    \begin{tabular}{|l|ccc|ccc|ccc|ccc|}
        \hline
        & \multicolumn{3}{c|}{Same-Text Dataset vs Dataset} & \multicolumn{3}{c|}{Random Dataset vs Dataset} & \multicolumn{3}{c|}{Generated vs Dataset} & \multicolumn{3}{c|}{Generated vs Generated} \\
        \cline{2-13}
        Metric & min & mean & std & min & mean & std & min & mean & std & min & mean & std \\
        \hline
        0/1 Hamming Distance & 0.00 & 96.11 & 74.16 & 16.00 & 177.97 & 62.23 & 133 & 214.12 & 26.76 & 124 & 237.61 & 56.27 \\
        \hline
        Euclidean Distance & 0.00 & 2.67 & 1.09 & 1.67 & 3.60 & 0.59 & 2.69 & 3.00 & 0.11 & 3.13 & 4.19 & 0.35 \\
        \hline
    \end{tabular}
    \caption{\small Comparison of Hamming and Euclidean distances across dataset and generated drumbeats. The first two columns are to provide a sense of scale. The first column values are for drumbeats from the dataset with the same text multihot vector (even though the original full text may be different). For the second column drumbeats are generated by randomly sampling 2 datapoints from the dataset. The third column focuses on {\em novelty} or {\em generalisation}, by comparing generated drumbeats with the closest 1\% of the dataset drumbeats. The last column compares multiple drumbeats generated from the same multihot text encoding. This demonstrates {\em variety} by highlighting that each drumbeat is unique.
    \label{tab:distance_metrics}}
\end{table*}

In our exploration of the model's behavior under perturbation, we explored the addition of noise into the latent variable $Z$ during training to enhance the robustness and stability of the latent space against perturbations. Our investigation focused on three distinct levels of noise application: High Noise, Low Noise, and No Noise. For High Noise, we introduced noise within the range [0.01, 0.1], and for Low Noise, the range was set between [0.01, 0.001]. The No Noise condition did not involve any additional noise. The findings indicated that the High Noise model yielded the most unique drumbeats, albeit at the expense of quality. In contrast, the drumbeats generated under Low and No Noise conditions were similar in terms of both quality and their proximity in the latent space.

\subsection{Listening Test}

To evaluate the quality of generated drumbeats and their correlation with text prompts, we constructed a survey comprising 40 drumbeats. These were created using 10 randomly selected text prompts from our dataset, resulting in four drumbeat variants for each prompt: (1) the original dataset drumbeat, (2) a drumbeat generated via the multihot text encoding, (3) a drumbeat generated by the BERT text encoding, and (4) a drumbeat generated with empty text, serving as a control or negative example. 


Drumbeats ranged from 8 to 15 seconds in length, leading to an estimated survey completion time of 30 to 60 minutes. Participants were blind to the specifics of the drumbeat generation methods along with the nature of each variant. There were three evaluation criteria: quality, aptness with the given text prompt, and novelty. These were each assessed using a Likert scale.


\begin{figure}
 \centerline{
 \includegraphics[width=1\columnwidth]{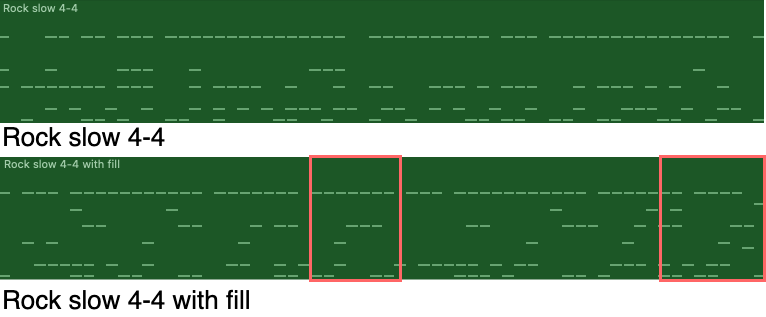}}
\caption{\small Two MIDI files with different text prompt shows that the MIDI file generated with \emph{``fill''} has fills added to it. Both the files can be heard by the readers on https://soundcloud.com/user-32049071/rock-slow-4-4 and https://soundcloud.com/user-32049071/rock-slow-4-4-with-fills}
\label{fig:midi_files_for_text}
\end{figure}

We received a total of 14 responses to our survey, out of which 12 were fully completed and used for all subsequent statistical evaluations. 

\begin{figure}
 \centerline{
 \includegraphics[width=1\columnwidth]{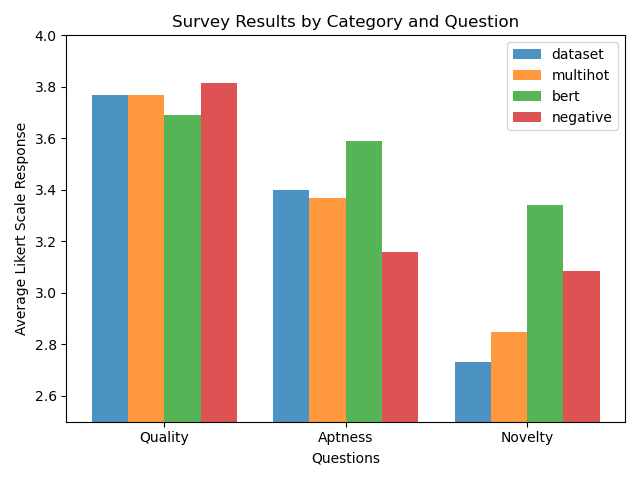}}
 \caption{\small The aggregated responses from N=12 participants of the survey rating each category of drumbeat on Quality, Suitability to the Text Prompt and How fresh or novel it was. The y-axis shows the values based on the Likert scale of 1-5, magnified between 2.5-4. The figure shows that the quality across all categories is roughly the same showcasing that the model is able to generate quality drumbeats. The BERT model performs the best in aptness to text and novelty closely followed by the dataset and multihot model.}
 \label{fig:survey_results}
\end{figure}

The results, depicted in Figure~\ref{fig:survey_results}, indicate that participants generally perceived the quality of each category of drumbeats as comparable and satisfactory. Thus, generated drumbeats were {\em as good as} professionally-recorded ones. 

Concerning `aptness', as hypothesized the generated drumbeats were much more apt to the text, versus the control drumbeats which were generated by ignoring the text.

The BERT model, in particular, was noted for its ability to generate drumbeats that were perceived as notably novel. It is also worth mentioning that the drumbeats originating from the dataset, which were created by human musicians, received the lowest scores for novelty.

\section{Conclusion and Future Work}\label{sec:conc}
In this study, we developed a system for generating drumbeats conditioned on textual prompts using Latent Diffusion Models. The results demonstrate the system's capability to produce high-quality, coherent drumbeats that aligns with human perception. Analysis from both the Listening Test and the empirical data in Table~\ref{tab:distance_metrics} confirms the system's capability to generate new drumbeats. Figures~\ref{fig:text_conditioning},~\ref{fig:midi_files_for_text},~and~\ref{fig:survey_results} illustrate that text conditioning plays an important role in generating the drumbeats. Participants in the listening test rated the quality of these drumbeats as comparable to those in \emph{The Groove Monkee} dataset, underscoring our model's ability to create drumbeats that match human-generated ones. The use of diffusion in the latent space enhances the quality and generation speed and opens possibilities for real-time integration. We have made the code along with some generated samples available for the readers on GitHub\footnote{https://github.com/pushkarjajoria/Text-Conditioned-Drumbeat-Generation} and SoundCloud\footnote{https://soundcloud.com/user-32049071/sets/generated-drumbeats}.

Looking ahead there are some improvements that could be made to the system. The text prompt generation could be improved by employing text augmentation techniques that converts the keyword type textual information into a more free flow natural language. LLMs models like BERT are better suited for free flow natural language and hence such augmentation may also result in better text embeddings. Conducting a larger-scale study in a more controlled environment will provide a more comprehensive data, providing insight into the human perceived capabilities of the system. The observed differences in Same-Text vs Different-Text in the Latent space warrants further investigation into how text prompts shape the musical output in the encoded or latent space. Investigating more by isolating the dimensions with high variance might yield some meaningful insights about the latent space and hence on the impact on text conditioning in the latent space. 

\section{Acknowledgments}
This work was carried out at the University of Galway with funding from the Polifonia Project of the European Union’s Horizon 2020 research and innovation programme under grant agreement N. 101004746.


\section{Ethics Statement}
The Groove Monkee Mega-Pack dataset is a commercial product available on \url{https://groovemonkee.com/products/mega-pack}. We train on it for research purposes only. We do not offer this dataset for download, nor do we offer our trained model for download as it could be seen as competing with Groove Monkee's commercial product. A free demo version of the dataset is available from Groove Monkee, enabling the reader to test our code.

\bibliography{DrumbeatGeneration}

\begin{thebibliography}{10}
\providecommand{\url}[1]{#1}
\csname url@samestyle\endcsname
\providecommand{\newblock}{\relax}
\providecommand{\bibinfo}[2]{#2}
\providecommand{\BIBentrySTDinterwordspacing}{\spaceskip=0pt\relax}
\providecommand{\BIBentryALTinterwordstretchfactor}{4}
\providecommand{\BIBentryALTinterwordspacing}{\spaceskip=\fontdimen2\font plus
\BIBentryALTinterwordstretchfactor\fontdimen3\font minus \fontdimen4\font\relax}
\providecommand{\BIBforeignlanguage}[2]{{%
\expandafter\ifx\csname l@#1\endcsname\relax
\typeout{** WARNING: IEEEtran.bst: No hyphenation pattern has been}%
\typeout{** loaded for the language `#1'. Using the pattern for}%
\typeout{** the default language instead.}%
\else
\language=\csname l@#1\endcsname
\fi
#2}}
\providecommand{\BIBdecl}{\relax}
\BIBdecl

\bibitem{biles2002genjam}
J.~A. Biles, ``Genjam: Evolution of a jazz improviser,'' in \emph{Creative evolutionary systems}.\hskip 1em plus 0.5em minus 0.4em\relax Elsevier, 2002, pp. 165--187.

\bibitem{eck2002first}
D.~Eck and J.~Schmidhuber, ``A first look at music composition using lstm recurrent neural networks,'' \emph{Istituto Dalle Molle Di Studi Sull Intelligenza Artificiale}, vol. 103, no.~4, pp. 48--56, 2002.

\bibitem{roberts2018hierarchical}
A.~Roberts, J.~Engel, C.~Raffel, C.~Hawthorne, and D.~Eck, ``A hierarchical latent vector model for learning long-term structure in music,'' in \emph{International conference on machine learning}.\hskip 1em plus 0.5em minus 0.4em\relax PMLR, 2018, pp. 4364--4373.

\bibitem{briot2019deep}
J.-P. Briot, G.~Hadjeres, and F.-D. Pachet, ``Deep learning techniques for music generation -- a survey,'' 2019.

\bibitem{Herremans_2017}
\BIBentryALTinterwordspacing
D.~Herremans, C.-H. Chuan, and E.~Chew, ``A functional taxonomy of music generation systems,'' \emph{ACM Computing Surveys}, vol.~50, no.~5, p. 1–30, Sep. 2017. [Online]. Available: \url{http://dx.doi.org/10.1145/3108242}
\BIBentrySTDinterwordspacing

\bibitem{Herremans_2019}
\BIBentryALTinterwordspacing
D.~Herremans and E.~Chew, ``Morpheus: Generating structured music with constrained patterns and tension,'' \emph{IEEE Transactions on Affective Computing}, vol.~10, no.~4, p. 510–523, Oct. 2019. [Online]. Available: \url{http://dx.doi.org/10.1109/TAFFC.2017.2737984}
\BIBentrySTDinterwordspacing

\bibitem{musical_creativity_Hewitt}
A.~HEWITT, ``Musical creativity: Multidisciplinary research in theory and practice edited by irène deliège and geraint wiggins. andover: Psychology press, 2006. 376 pp, £60.00, hardback. isbn: 1-84169-508-4,'' \emph{British Journal of Music Education}, vol.~24, no.~3, p. 343–346, 2007.

\bibitem{dahale2022generating}
R.~Dahale, V.~Talwadker, P.~Rao, and P.~Verma, ``Generating coherent drum accompaniment with fills and improvisations,'' 2022.

\bibitem{huang2018music}
C.-Z.~A. Huang, A.~Vaswani, J.~Uszkoreit, N.~Shazeer, I.~Simon, C.~Hawthorne, A.~M. Dai, M.~D. Hoffman, M.~Dinculescu, and D.~Eck, ``Music transformer,'' 2018.

\bibitem{radford2021learning}
A.~Radford, J.~W. Kim, C.~Hallacy, A.~Ramesh, G.~Goh, S.~Agarwal, G.~Sastry, A.~Askell, P.~Mishkin, J.~Clark, G.~Krueger, and I.~Sutskever, ``Learning transferable visual models from natural language supervision,'' 2021.

\bibitem{ho2020denoising}
J.~Ho, A.~Jain, and P.~Abbeel, ``Denoising diffusion probabilistic models,'' 2020.

\bibitem{nichol2021improved}
A.~Nichol and P.~Dhariwal, ``Improved denoising diffusion probabilistic models,'' 2021.

\bibitem{rombach2022highresolution}
R.~Rombach, A.~Blattmann, D.~Lorenz, P.~Esser, and B.~Ommer, ``High-resolution image synthesis with latent diffusion models,'' 2022.

\bibitem{MontesinosLópez2022}
\BIBentryALTinterwordspacing
O.~A. Montesinos~L{\'o}pez, A.~Montesinos~L{\'o}pez, and J.~Crossa, \emph{Convolutional Neural Networks}.\hskip 1em plus 0.5em minus 0.4em\relax Cham: Springer International Publishing, 2022, pp. 533--577. [Online]. Available: \url{https://doi.org/10.1007/978-3-030-89010-0_13}
\BIBentrySTDinterwordspacing

\bibitem{McDermott2021}
\BIBentryALTinterwordspacing
J.~McDermott, \emph{Representation Learning for the Arts: A Case Study Using Variational Autoencoders for Drum Loops}.\hskip 1em plus 0.5em minus 0.4em\relax Cham: Springer International Publishing, 2021, pp. 139--161. [Online]. Available: \url{https://doi.org/10.1007/978-3-030-59475-6_6}
\BIBentrySTDinterwordspacing

\bibitem{Tikhonov}
A.~Tikhonov and I.~Yamshchikov, ``Drum beats and where to find them: Sampling drum patterns from a latent space,'' 07 2020.

\bibitem{makris2022conditional}
D.~Makris, G.~Zixun, M.~Kaliakatsos-Papakostas, and D.~Herremans, ``Conditional drums generation using compound word representations,'' 2022.

\bibitem{wu2022jukedrummer}
Y.-K. Wu, C.-Y. Chiu, and Y.-H. Yang, ``Jukedrummer: Conditional beat-aware audio-domain drum accompaniment generation via transformer vq-vae,'' 2022.

\bibitem{kaliakatsos2018generating}
M.~Kaliakatsos-Papakostas, ``Generating drum rhythms through data-driven conceptual blending of features and genetic algorithms,'' in \emph{Computational Intelligence in Music, Sound, Art and Design: 7th International Conference, EvoMUSART 2018, Parma, Italy, April 4-6, 2018, Proceedings}.\hskip 1em plus 0.5em minus 0.4em\relax Springer, 2018, pp. 145--160.

\bibitem{hoover2008scaffolding}
A.~K. Hoover, M.~P. Rosario, and K.~O. Stanley, ``Scaffolding for interactively evolving novel drum tracks for existing songs,'' in \emph{Applications of Evolutionary Computing: EvoWorkshops 2008: EvoCOMNET, EvoFIN, EvoHOT, EvoIASP, EvoMUSART, EvoNUM, EvoSTOC, and EvoTransLog, Naples, Italy, March 26-28, 2008. Proceedings}.\hskip 1em plus 0.5em minus 0.4em\relax Springer, 2008, pp. 412--422.

\bibitem{roberts2019hierarchical}
A.~Roberts, J.~Engel, C.~Raffel, C.~Hawthorne, and D.~Eck, ``A hierarchical latent vector model for learning long-term structure in music,'' 2019.

\bibitem{Blankensmith2018}
\BIBentryALTinterwordspacing
T.~Blankensmith and K.~Phillips, ``Beat blender,'' 2018. [Online]. Available: \url{https://experiments.withgoogle.com/ai/beat-blender/view/}
\BIBentrySTDinterwordspacing

\bibitem{callender2020improving}
L.~Callender, C.~Hawthorne, and J.~Engel, ``Improving perceptual quality of drum transcription with the expanded groove midi dataset,'' 2020.

\bibitem{wu2023clamp}
S.~Wu, D.~Yu, X.~Tan, and M.~Sun, ``Clamp: Contrastive language-music pre-training for cross-modal symbolic music information retrieval,'' 2023.

\bibitem{6795963}
S.~Hochreiter and J.~Schmidhuber, ``Long short-term memory,'' \emph{Neural Computation}, vol.~9, no.~8, pp. 1735--1780, 1997.

\bibitem{yang2020evaluation}
L.-C. Yang and A.~Lerch, ``On the evaluation of generative models in music,'' \emph{Neural Computing and Applications}, vol.~32, no.~9, pp. 4773--4784, 2020.

\end{thebibliography}

\end{document}